\documentstyle[pre,aps,multicol,epsf]{revtex} 
\begin{document}
\title{{\hfill} {\sl Submitted}\\ 
Extreme Wave Events in Directional, Random Oceanic Sea States 
}
\author{M. Onorato, A. R. Osborne and M. Serio
}
\address{
Dipartimento di Fisica Generale, Universit\'{a} di Torino,
Torino, Via P. Giuria, 1 - 10125, ITALY
}
\maketitle
\begin{abstract} 
We discuss the effects of the
directional spreading on the occurrence of extreme wave events. We numerically
integrate the envelope equation recently proposed by Trulsen et al., Phys of
Fluids 2000, as a weakly nonlinear model for realistic oceanic gravity waves.
Initial conditions for numerical simulations are characterized by the spatial
JONSWAP power spectrum for several values of the significant wave height,
steepness and directional spreading. We show that by increasing the
directionality of the initial spectrum the appearance of extreme events is
notably reduced. \end{abstract}
%

\begin{multicols}{2} 

We address the occurrence of extreme wave events in numerical models
of random, directional, oceanic sea states. Extreme ocean waves of this type are
of unusually large size with respect to the background, surrounding waves. They
are often refered to  as ``freak'' or ``rogue'' waves and have been given a
rough definition in terms of an arbitrary threshold, $H_{\max }>2.2H_{s},$ for
$H_{s}$ the significant wave height \cite{DEAN}. 
Because extreme waves are rare, and hence are not often
measured , the physical mechanisms for their occurrence have not been
unequivocally established. Possible causes of freak waves in the open sea have
been investigated by many researchers \cite{BRE} and basically three mechanisms 
have been proposed: 
the linear interaction of the waves with the currents 
(geometrical optic theory), see \cite{PER1,LAV,WFO};
 the simple linear superposition of Fourier phases 
(the resulting large waves are also known 
as transient waves \cite{KRIBLE}) and the modulation instability \cite{TDF}.

The last mechanism and indeed the method addressed 
herein is based on the nonlinear phenomenon known as
the Benjamin-Feir instability \cite{BFI}, which describes how a monochromatic
wave train can be unstable to small, side-band perturbations. This physical
phenomena has been widely studied in wave tank facilities (see for example
\cite{YUEL}, \cite{TUL} and references therein) and from a theoretical and
numerical point of view for the 1+1 Nonlinear Schroedinger equation (NLS
equation in one space and one time dimensions) \cite{TRA,YUL} and for the fully
nonlinear Euler equations \cite {HPD}. A major complication arises from the fact
the envelope equations (for example NLS) are
derived from the primitive equations of motion under the hypothesis of a
narrow-banded spectrum. Higher order equations in the envelope hierarchy have
then been proposed \cite{DYS,TDE} in order to overcome the limitation of narrow
bandwidth. Numerical simulations of these equations in 1+1 dimension \cite
{ONBRE,ONPRL} have shown that the probability of occurrence of freak waves is
strictly related to the ``enhancement'' coefficient $\gamma $ and the Phillips
parameter $\alpha $ of the JONSWAP power spectrum \cite{KOM}. In
\cite{ONBRE,ONPRL} it was found that increasing the values of $\gamma $ and
$\alpha $ has the effect of increasing the probability of finding freak wave
events. 
In addition to the physical limitations associated with the narrow spectral width
assumption, an even more severe limitation of the above results arises because
these conclusions have been obtained from essentially one-dimensional numerical
simulations.

In this communication we address the less restricted problem of the appearance
of freak waves in 2+1 dimensions, i.e. the wave field depends on the spatial
coordinates $x$, $y$ and time, $t$, so that directional spreading is
systematically included. Preliminary
results have already been considered in \cite{OSB00} where simple initial
conditions characterized by a carrier wave plus longitudinal and transverse
perturbations showed an ubiquitous formation of large amplitude ''freak'' waves
under the evolution of the NLS equation in 2+1.
Our aim here, instead, is to study the more realistic case characterized by the
JONSWAP spectrum for a directional wave train, while at the same time extending
the order of the simulations so that energy leakage to high wave numbers does
not occur (see \cite{YUEL}). Recently Trulsen et al. \cite{TKDV} proposed
a model equation (see eqs. (\ref{DLM1}-\ref{DLM2}) below)
that shares many of the
properties of the Zakharov equation  \cite{ZAKH}.
The main advantage of the model is that it 
is numerically inexpensive and therefore
suitable for extensive numerical simulations
in which large amounts of
statistical information is desired.
The equation can be thought of as a special
case of the Zakharov equation:  
\begin{eqnarray} 
&& i \frac {\partial B({\bf k},t)} {\partial t} = 
\int_{-\infty}^{+\infty}
T_{{\bf k}, {\bf k}_1, {\bf k}_2, {\bf k}_3} 
B({\bf k}_1,t)^* B({\bf k}_2,t) B({\bf k}_3,t) \times \nonumber\\
&& \delta({\bf k} +{\bf k}_1-{\bf k}_2-{\bf k}_3)
e^{i(\omega+\omega_1-\omega_2-\omega_3)t}
d{\bf k}_1d{\bf k}_2d{\bf k}_3,
\label{zakharov}
\end{eqnarray}
where $T_{{\bf k}, {\bf k}_1, {\bf k}_2, {\bf k}_3}=
T({\bf k}, {\bf k}_1, {\bf k}_2, {\bf k}_3)$ is the coupling coefficient 
(for the analytical form of it see for example \cite{KRA}).
We briefly outline the connection of the Zakharov equation with the 
proposed model. One performs the following change of variables: ${\bf
k=k_{0}+\bbox{\chi}}$ with $ \bbox{\chi}=(\lambda ,\mu )$ and  
\begin{equation} {
A( \bbox{\chi},t)=B({\bf k},t) e^{-i(\omega({\bf k}) -
\omega({\bf k_0}))t}
}\label{changevar2}
\end{equation}
For simplicity we perform a simple rotation so
that ${\bf k}_{0}$ has the coordinates $(k_{0},0)$. The next step consists in a
Taylor's expansion of the coupling coefficient, $T({\bf k},{\bf k}_{1},{\bf
k}_{2},{\bf k}_{3}),$ up to first order in spectral width around the carrier
wave number, ${\bf  k}_{0}$. Note that the expansion is performed on the
nonlinear term, while leaving the linear part of the equation unchanged. As a
result, the following particular kernel is used in eq. (\ref{zakharov}) (see
\cite{STI} ): 
\begin{eqnarray} 
T( {\bf k_0} + \bbox{\chi}, {\bf k_0}+\bbox{\chi}_1,
{\bf k_0}+\bbox{\chi}_2, {\bf k_0}+\bbox{\chi}_3)  = 
\frac{k_0^3} {4 \pi^2} \bigg(1+  \nonumber\\ 
\frac{3} {2k_0} (\lambda_2+\lambda_3)-
\frac{(\lambda_1-\lambda_2)^2} {2k_0|\bbox{\chi}_1-\bbox{\chi}_2|} - 
\frac{(\lambda_1-\lambda_3)^2} {2k_0|\bbox{\chi}_1-\bbox{\chi}_3|} \bigg)
\label{kernel2}
\end{eqnarray}
Under these conditions the triple integral in eq. (\ref{zakharov})
can be performed (see \cite{STI} for details) and the following set of equations
is obtained in physical (configuration) space: 
\begin{eqnarray} 
&&  A_t+ \omega_0 L(\partial_x,\partial_y)A+
i \frac{\omega_0 k_0^2} {2} \mid A\mid^2 A +
\frac{3\omega_0 k_0} {2} \mid A\mid^2 A_{x} - \nonumber\\ 
&& \frac{\omega_0k_0} {4} A^2 A_{x}^*
+i k_0 A\bar\phi_x|_{z=0}=0
 \label{DLM1}
\end{eqnarray}
where $L=i\left[ \left( (1-i\partial _{x})^{2}-i(\partial _{y})^{2}\right)
^{1/4}-1\right]$; use of this latter operator insures retention of linear
dispersion to all orders. $\bar\phi$ is the inducead mean flow which is coupled with the envelope
$A$ via the following nonlocal relation:
\begin{equation}
{\bar\phi_x|_{z=0}=-\frac{\omega_0} {2}F^{-1}\{|k|F\{|A|^2\}\}
,} \label{DLM2}
\end{equation}
with $F$ the Fourier operator. We call 
equations (\ref{DLM1}-\ref{DLM2}) the
Generalized Dysthe Equation (GDE) since the nonlinear part is exactly that of
the Dysthe Equation \cite{DYS}, but further includes the linear dispersion
relation of the inviscid primitive equations of motion. As stated in
\cite{TKDV}, this equation shares many properties of the Zakharov equation:
the four wave resonant interaction curve coincides with the Phillips
curve and moreover the instability region is bounded and is quantitatively
similar to the results obtained by McLean \cite{MCL} for the Zakharov equation
(see Fig. 1 in \cite{TKDV}).

Since our main objective herein relates to the generation of freak waves by the
Benjamin-Feir instability and since eqs. (\ref{DLM1}-\ref{DLM2}) show good
agreement with the fully nonlinear inviscid equations for the two dimensional
instability diagram, we feel that many of the features of the primitive
equations are retained in the GDE equation. This suggests that a study of the
generation of freak waves using this higher order formulation may indeed allow
us to investigate many of the effects of directional spreading on the generation
of freak waves.

The GDE equations have been solved numerically using a standard pseudo-spectral
code fully dealiased and with second order time-stepping on a doubly-periodic
domain with a spatial discretization of $256\times 64$. The initial conditions
consist of a two-dimensional wave field characterized by the JONSWAP power
spectrum. In order to statistically search for extreme wave events we focus on
the temporal evolution of the fourth-order, central moment ({\it kurtosis}) of
the probability density function of the wave amplitude.
The kurtosis is a statistical measure of the probability density function of
the wave amplitudes:  the larger/wider the tails are, the larger are the value
of the kurtosis. It is clear that when the kurtosis is much larger than 3 (the
well-known value for a Gaussian process)  we are faced, in physical space, with
an extreme wave event that likely corresponds to a ''freak'' wave (see below for
details).

\begin{figure} 
\epsfxsize=8.5cm \epsfbox{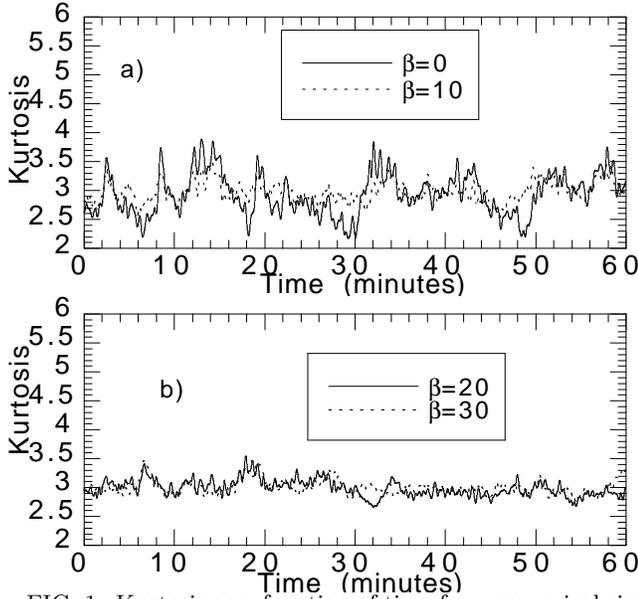}
\caption{Kurtosis as a function of time from numerical
simulations with $ \alpha=0.0081$ and $\gamma=1$, for $\beta$=0,10 (a) and
$\beta$=20,30 (b) .} 
\label{fig kurto1} \end{figure}

\begin{figure} 
\epsfxsize=8.5cm \epsfbox{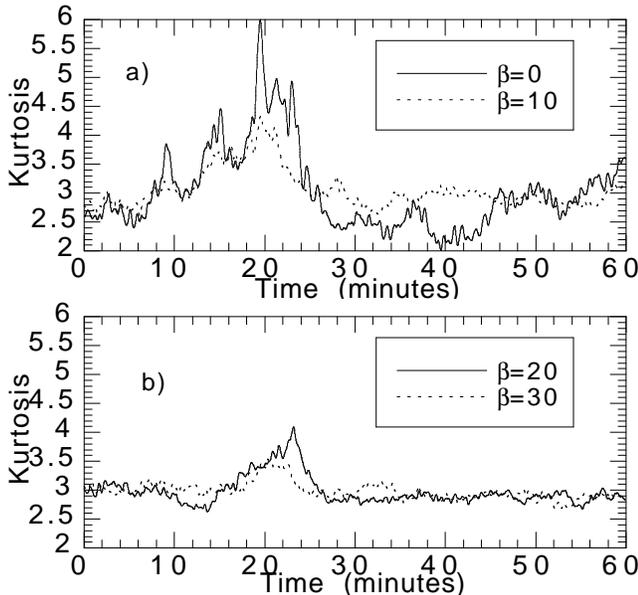}
\caption{Kurtosis as a function of time from numerical
simulations with $ \alpha=0.015$ and $\gamma=5$, for $\beta$=0,10 (a) and
$\beta$=20,30 (b) .} 
\label{fig kurto2}
\end{figure}

\begin{figure} 
\epsfxsize=8.5cm \epsfbox{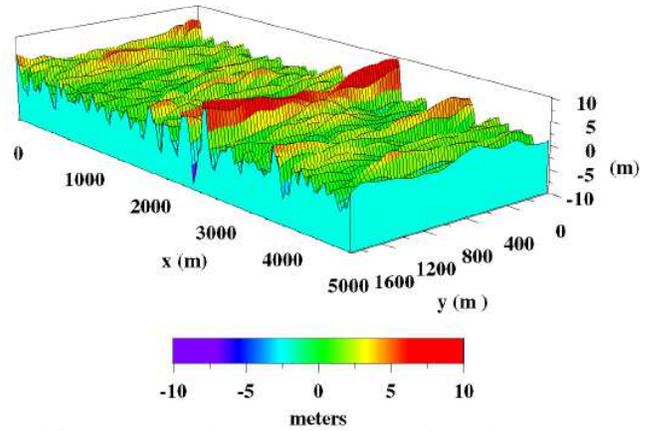}
\caption{A two dimensional free surface showing a quasi gaussian backgraound 
and a large wave amplitude inthe middle of it.
The kurtosis of the field is 4.3.
The surface has been obtained from the nonlinear evolution of a 
random phase wave field characterized by 
a JONSWAP spectrum with $\alpha=0.0081$, $\gamma=5$ 
and $\beta=10$. } \label{fig freak}
\end{figure}

In order to prepare the initial wave field it is necessary to generate a
directional, frequency spectrum $S(f,\theta )=P(f)G(\theta )$, where $P(f)$ is
the original JONSWAP spectrum, and then to transform it into the associated wave
number spectrum, $S(k_{x},k_{y})$. The directional spreading function $G(\theta
)$ used here is a cosine-squared function in which only the first lobe (relative
to the dominant wave direction) is considered:  
\begin{equation}
{G(\theta)=\left\{  
\begin{array}{ll}
\cos^2 \bigg( \frac{\pi} {2 \beta} \theta \bigg)  
 & if -\beta \leq \theta \leq \beta  \\
0 & else
\end{array}
\right. 
} \label{spreading}
\end{equation}
$\beta $ is an in important parameter that
provides a measure of the directional spreading, i.e. as $\beta \rightarrow 0$
the waves become increasingly unidirectional. Using the linear dispersion
relation for infinite water depth, $\omega =\sqrt{g|{\bf k}|}$, the wave number
spectrum can be written:  
\begin{equation} {
S(k_x,k_y)=\frac {\alpha} {|\bf {k}|^3} 
\frac{1} {2 |\bf{k}|} 
e^ {-\frac{3} {2} \left(\frac{k_0} {|\bf{k}|}\right)^2 } 
\gamma^
{\exp \bigg[-\frac{ \left(\sqrt{|\bf{k}|}-\sqrt{k_0}\right)^2} 
{2\sigma_0^2 k_0^2}\bigg] } G(\theta)
} \label{2Dspectrum}
\end{equation}
where $\theta =\arctan (k_{y}/k_{x})$. From
eq. (\ref{2Dspectrum}) the two-dimensional surface elevation is computed in the
following simple way (see \cite{BOR}): 
\begin{equation} 
{\eta(x,y)=\sum_{i=1}^{N} \sum_{j=1}^{M} C_{ij} \cos(k_i x +k_j y
+\phi_{ij}), } \label{rsurface}
\end{equation}
where the $\phi _{ij}$ are uniformly distributed
random phases on the interval $(0,2\pi )$ and
$C_{ij}=\sqrt{4S(k_{i},k_{j})\Delta k_{i}\Delta  k_{j}}$.
Most of our numerical simulations have been 
started with a field of size $5026.5$ m $\times $
$1885$ m, with a dominant wave number $ k_{0}=0.02$ m$^{-1},$ corresponding to a
characteristic wavelength of $\sim 300$ m. The time step
adopted is 1 second and the full duration of the simulation for each run is one
hour in real time. We recall that the nonlinear time scale $T_{NL}$ for the
effects of the Benjamin-Feir instability to appear is of the order of
$O(1/(\varepsilon ^{2}\omega _{0}))$. In our numerical simulations the wave
steepness is $\varepsilon \sim 0.1$ and the dominant frequency $f_{0}$ can be
estimated from $k_{0}$ using the linear dispersion relation in infinite water
depth, $f_{0}\simeq 0.07$ s$^{-1}$. A rough estimation of the nonlinear time
scale is $T_{NL}\simeq 250$ s. For each time step the two-dimensional surface
has been visualized and the kurtosis has been computed from the space series
associated with each instant in time of the complete evolution. 
In Fig. \ref{fig kurto1}a, b we
show that, for small values of the enhancement parameter $\gamma $ (for $\gamma =1$
the significant wave height
$H_{s}$, computed as 4 times the standard deviation of the wave amplitude, was
estimated to be $H_{s}=7.26$ m ), the kurtosis is found to be
$\simeq 3$ for all the evolution, 
independently of the value of the spreading parameter $\beta$. Additional
numerical simulations, with many different sets of random phases, have been
performed and the results are quantitatively similar to the one just presented.
A more interesting case, from the point of view of extreme
waves, corresponds to the numerical simulations with initial conditions
characterized by $\gamma =5$ ($H_{s}=9.8$ m). The main results of these latter
simulations are shown in Fig. \ref{fig kurto2}a, b. 
%
Here we see that for small
values of the spreading parameter, $\beta $, the kurtosis suddenly starts to
increase (Fig. \ref{fig kurto2}a) indicating the {\it onset} of the
Benjamin-Fier instability. The kurtosis is seen to grow, to reach a maximum
value and then to decrease and oscillate around the value $\simeq 3$. An
example of the wave field with the $kurtosis=4.3$,  resulting from the evolution
of an initial condition with $\gamma=5$ and $\beta=10$ is given in 
Fig. \ref{fig freak}: a large amplitude wave with respect to the 
rest of the quasi-gaussian background appears in the middle of the
two dimensional surface plot; note also that before the large 
wave a deep hole is also evident (this phenomenology 
is consistent with many descriptions of mariners who happened to run into 
a freak wave). As the
spreading parameter, $\beta $, is increased in subsequent numerical simulations
(Fig. \ref{fig kurto2}b) the kurtosis continues to grow but does not reach
values as high as in the previous cases. For the larger values of $\beta $, the
kurtosis oscillates around $\simeq 3$, indicating that the probability
density function for the wave amplitudes is approximately Gaussian.
A considerable number of numerical simulations have been carried out using
different values for the selected sets of random phases. The results are in
qualitative agreement with the case just discussed. However, we find that the
times at which the rare events appear are different for different sets of random
phases and that the maximum values of the kurtosis can be also be quite
different. For a number of random realizations we have found, for the case of a
quasi-unidirectional spectrum, the appearance of as many as two extreme events
in a single hour of simulation time. We note that for very small directional
spreading (essentially the one-dimensional case) the pattern of freak wave
generation is almost periodic, primarily because of the well-known phenomenon of
Fermi-Pasta-Ulam recurrence. In the fully 2+1 case the physics is much more
complicated: as noted in \cite{TDD}, there is a permanent frequency downshift
which excludes the possibility of recurrence.
As a final general consideration it should be mentioned that the classical  {\it
statistical description} of deep water wave trains (using the kinetic 
equations of the Hasselmann, or Boltzmann, 
type in the quasi-Gaussian approximation) is based on
the assumption that the wave field is spatially homogeneous (see for example
\cite{KOM}): under this condition the Benjamin-Feir instability is suppressed
and the transfer of energy in the power spectrum is ruled only by the {\it
four-wave resonant interaction} whose nonlinear time scale is 
$T_{NL}=O(1/(\varepsilon ^{4}f_{0}))$. Therefore, if the 
statiscal prediction of the occurence of freak waves is desired, a new form of 
kinetic equation should be adopted. This very interesting application is,
however, beyond the scope of the present paper and will be considered in a
future work.

{\bf Acknowledgements}
The authors are grateful to D. Resio for valuable discussions
at the beginning of the work.
M. O. was supported by a Research Contract from the Universit\`{a} di
Torino. This work was supported by  the Office of Naval Research of the
United States of America (T. F. Swean, Jr. and T. Kinder) and by the Mobile
Offshore Base Program of ONR (G. Remmers). Consortium funds and Torino
University funds (60 \%) are also acknowledged.
%
%

\end{multicols} 

\end{document}